\documentclass[prb, twocolumn]{revtex4-1}

\usepackage{amsmath}
\usepackage{graphicx}
\usepackage{subfigure}

\begin{document}

\title{Kinetic exchange models: From molecular physics to social science}

\author{Marco Patriarca}
\email{marco.patriarca@kbfi.ee}
\affiliation{National Institute of Chemical Physics and Biophysics, R\"avala 10, Tallinn 15042, Estonia}

\author{Anirban Chakraborti}
\email{anirban.chakraborti@ecp.fr}
\affiliation{Laboratoire de Math\'{e}matiques Appliqu\'{e}es aux Syst\`{e}mes, \'{E}cole Centrale Paris, 92290 Ch\^{a}tenay-Malabry, France}

\date{\today}

\begin{abstract}
We discuss several multi-agent models that have
their origin in the kinetic exchange theory of statistical mechanics and have been recently applied to a variety of problems in the social sciences. This class of models can be easily adapted for simulations in areas other than physics, such as the modeling of income and wealth distributions in economics and opinion dynamics in sociology.
\end{abstract}

\maketitle

\section{Introduction}

The application of probability theory to the study of gases led to the
formulation of the kinetic theory of gases, prepared the
basis for the formulation of the Maxwell velocity distribution, and later, for the development of statistical
mechanics.\cite{Boltzmann_A,terHaar1995,Sethna_A} The initial triumph was that the empirical laws of thermodynamics were obtained from the basic assumptions of statistical mechanics. Today we understand that the scope of statistical mechanics is much broader. Because it has the tools to treat macroscopic systems with a large number of microscopic constituents, it can be naturally applied to many types of systems.

Kinetic exchange models, the subject of this
article, are stochastic models which have a straightforward interpretation in terms of energy exchanges in a gas.
However, they can be suitably adapted and used to
study problems in the social sciences, as illustrated by recent work.\cite{Patriarca2010b,Chatterjee2007b,Chakraborti2011b,Lallouache2010a}
The applications of kinetic exchange models to problems in fields
from molecular physics to the social sciences illustrates the historical link between statistical mechanics and the social sciences. Recently, this link has been rediscovered due to the development of complex systems theory
and social dynamics.
In fact, statistics, which was an important basis
for Maxwell's and Boltzmann's work and the foundation of statistical
mechanics, originated from the study of demographic data.\cite{Ball2002a}
The idea that a large number of molecules and a large social group
have many important common features, in particular, that they are
predictable due to the high number of their components despite their intrinsic random character,
was shared by many investigators. Boltzmann wrote that
``molecules are like so many individuals, having the most various states of motion,'' when writing about the foundations of
statistical mechanics.\cite{Ball2002a,Boltzmann1872a}

Multi-agent models are a class of models where the actions and interactions of autonomous agents, which may represent individuals, organizations, societies, etc., can be used to understand the behavior of the system as a whole. 
The simple formulation and numerical
implementation of kinetic exchange models make them attractive in many 
disciplines.
They can be regarded as minimal prototypical models of complex systems consisting
of a set of (possibly) heterogeneous units interacting according to simple laws, yet they are able to exhibit well defined states with robust probability distributions.

In this article we 
focus on the formulation, interpretation, and 
simulation of a few representative examples of kinetic exchange models.
Other social science applications of kinetic exchange models are also summarized.

\section{\label{homogeneous}Homogeneous Kinetic Exchange Models}

\subsection{\label{molecular}Kinetic exchange models in molecular physics: Thermal relaxation in $\boldsymbol{d}$-dimensions}

We introduce the general structure of a kinetic exchange model by a
simple example. It is assumed that the $N$ (minimally) interacting units $\{i\}$, with $i=1,2,\dots,N$, are molecules of a gas with no interaction energy 
and the variables $\{w_i\}$ represent their kinetic energies, such that $w_i \ge 0$.
The time evolution of the system proceeds by a discrete stochastic dynamics. A series of updates
of the kinetic energies $w_i(t)$ are made at the discrete times $t = 0, 1, \dots$\,.
Each update takes into account the
effect of a collision between two molecules.
The time step, which can be set to $\Delta t = 1$ without loss of
generality, represents the average time interval between two
consecutive molecular collisions; that is, on average, after each time step
$\Delta t$, two molecules $i$ and $j$ undergo a scattering process and
an update of their kinetic energies $w_i$ and $w_j$ is made.

The evolution of the system is accomplished by the following steps at each time $t$.

\begin{enumerate}

\item Randomly choose a pair of molecules $i$ and $j$ ($i \ne j$ and $1 \le i,j \le N$), with kinetic energies $w_i$ and $w_j$, respectively; they represent the molecules undergoing a collision.

\item Compute the net amount $\Delta w_{ij}$ of kinetic energy to be exchanged between the molecules; its value is a function of the initial kinetic energies $w_i$ and $w_j$ and depends on the model considered.

\item Perform the energy exchange between $i$ and $j$ by updating their kinetic energies,
\begin{subequations}
\label{exchange}
\begin{align}
w_i &\to w_i - \Delta w_{ij} \\
w_j & \to w_j + \Delta w_{ij} \, .
\end{align}
\end{subequations}
The total kinetic energy is conserved during an interaction.

\item Set $t \to t + 1$ and go to step 1.

\end{enumerate}

The form of the function $\Delta w_{ij}$ defines the
specific model.
Kinetic exchange models describe the dynamics at a microscopic level, 
based on single molecular collisions. 
Such a representation can be optimal in terms of simplicity and
computational efficiency when the focus is on the energy dynamics, because
particles are described by their energy degree of freedom $w$ only, 
rather than by the entire set of their 2$d$ position and momentum coordinates, for a $d$-dimensional system. 

As our first simple example, consider the reshuffling rule: $w_i \to
\epsilon(w_i + w_j)$, $w_j \to (1 -\epsilon)(w_i + w_j)$, where
$\epsilon$ is a stochastic variable drawn as a uniform random number between 0 and 1. This rule corresponds
to $\Delta w_{ij} = (1 - \epsilon)w_i - \epsilon w_j$.
In this case, the algorithm we have outlined leads from arbitrary
initial conditions to the Boltzmann-Gibbs energy distribution at equilibrium, 
$f(w) = \beta \exp(-\beta w)$, where $\beta = 1/\langle w\rangle$ and $\langle w\rangle$ represents the mean energy of a single molecule. The theoretical derivations of this result using the Boltzmann transport equation, or entropy maximization principle, or simple probabilistic arguments, can be found in most standard textbooks of statistical mechanics.

As a more general example, consider the relaxation in energy
space of a gas in $d$-dimensions.
We assume that $d >1$ because the momentum and energy distributions of a
one-dimensional gas (where only head-on collisions occur)
do not change with time.
Although the model can be conceived most easily for a gas
in three dimensions, the most surprising features of kinetic exchange
models appear for a gas in an arbitrary number of dimensions, 
a case relevant for the treatment of heterogeneous systems in Sec.~\ref{heterogeneous}.
For a gas in $d$-dimensions the form of the update rule, and in particular of $\Delta w_{ij}$, 
can be derived exactly from energy and momentum conservation during a
collision between two particles $i$ and $j$.
If the respective $d$-dimensional vectors of the particle initial momenta are
$\mathbf{p}_i$ and $\mathbf{p}_j$, we find\cite{Chakraborti2008a}
\begin{align}
\label{deltax}
\Delta w_{ij} &= r_i w_i - r_j w_j \\
\label{rn}
r_k &= \cos^2 \alpha_k \quad (k = i, j) \\
\label{cosa}
\cos \alpha_k &= \frac{\mathbf{p}_k \cdot \Delta\mathbf{p}_{ij}} {|\mathbf{p}_k|\,|\Delta\mathbf{p}_{ij}|} \, ,
\end{align}
where $\cos \alpha_k$ is the direction cosine of momentum
$\mathbf{p}_k$ ($k = i, j$) with respect to the direction of the
transferred momentum $\Delta\mathbf{p}_{ij} = \mathbf{p}_i - \mathbf{p}_j$.
The directions of the two colliding particles can be assumed to be
random using the hypothesis of molecular chaos.\cite{terHaar1995}

We can now study the time evolution by randomly choosing at each time
step two new values for $r_k$ in Eq.~(\ref{deltax}), instead of maintaining a
list of momentum coordinates, as is commonly done in a molecular dynamics simulation.
Then we use Eq.~(\ref{exchange}) to compute the new particle
energies $w_i$ and $w_j$.
Note that the $r_k$'s are not uniformly distributed in
$(0,1)$, and thus the form of
their probability distribution function has to be chosen with some prudence.
In fact, their distribution strongly depends on the spatial dimension $d$,
their average value being $\langle r_k \rangle = 1/d$;�
see Ref.~\onlinecite{Chakraborti2008a} for further details.
This dependence of $\langle r_k \rangle$ on $d$ can be intuitively understood from kinetic theory: the greater the value of $d$, the more unlikely it becomes that $r_k$ assumes values close to
$r_k = 1$ (corresponding to a one-dimensional-like head on collision).�
Hence, a simple and computationally efficient choice�
is a uniform random distribution $f(r_k)$, limited in
the interval $(0, 2/d)$, 
such that the average value $\langle r_k \rangle = 1/d$.

Simulations of this model system 
using random numbers 
in place of the $r_i$'s in Eq.~(\ref{deltax}), for $d = 2$,
give the equilibrium Boltzmann-Gibbs distribution: $f(w) = \beta \exp(-\beta w)$, where $\beta = 1/\langle w\rangle $, as before. For $d > 2$, we obtain the $d$-dimensional
generalization of the standard Boltzmann
distribution,\cite{Patriarca2004a,Patriarca2004b,Chakraborti2008a}
namely the Gamma ($\Gamma$) distribution\cite{Abramowitz1970a,GammaDistribution1}
characterized by a shape parameter $\alpha$ equal to half of the spatial dimension,
\begin{align}
\label{gamma-a}
f(w, \alpha, \theta) &= \frac {w^{\alpha-1} e^{- w/\theta}} {\theta^\alpha \Gamma(\alpha)} \\
\label{gamma-b}
\alpha &= d / 2 \\
\label{gamma-c}
\theta &= \langle w \rangle / \alpha \, .
\end{align}
The scale parameter $\theta$ of the $\Gamma$-distribution is fixed, by
definition, by Eq.~(\ref{gamma-c}).\cite{Abramowitz1970a,GammaDistribution1}
From the equipartition theorem in classical statistical mechanics, $w = d \, k_\mathrm{B} T / 2$. 
Hence, we see that Eq.~(\ref{gamma-c}) identifies the scale parameter $\theta$ as the
absolute temperature (in energy units) given by $\theta \equiv k_\mathrm{B} T = 1/ \beta$.
Therefore, the same Boltzmann factor, $\exp(-w/\theta)$, is present in
the equilibrium distribution independently of the dimension $d$, and
the prefactor $w^{\alpha - 1}$ depends on $d$, because
it takes into account the phase-space volume proportional to $p^d \propto w^{d/2}$, where $p$ is the momentum
modulus.

\subsection{\label{social}Modeling the wealth distribution}

\begin{figure}[t]
\includegraphics*[width=8.5cm]{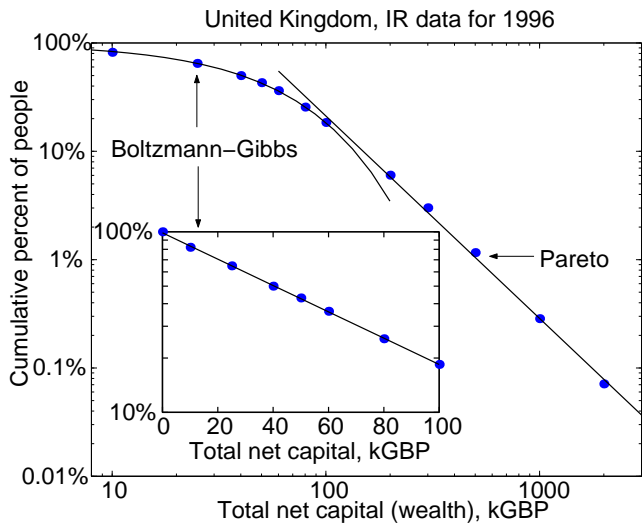}
\vspace{-0.25cm}
\caption{Cumulative probability distribution of the net wealth, composed of assets (including cash, stocks, property, and household goods) and liabilities (including mortgages and other debts) in the United Kingdom shown on log-log (main panel) and log-linear (inset) scales. Points represent the data from the Inland Revenue, and solid lines are fits to the Boltzmann-Gibbs (exponential) and Pareto (power) distributions.\cite{Yakovenko2001}
\label{fig_money2001}}
\end{figure}

Besides their interpretation in the context of physics, 
kinetic exchange models are applicable to problems in the
social sciences such as opinion dynamics and the modeling of
economic systems.
The latter models demonstrate their 
adaptability and promise,
particularly in their application to wealth distributions. In this section we consider how some kinetic exchange models describe the formation of the wealth distribution observed in
a society, as a consequence of binary wealth exchanges between
individuals.
We assume here that the standard definition of wealth is
the set of all those things with some monetary or exchange value. 
The shape of a typical wealth distribution is complex, 
with a Boltzmann-Gibbs (exponential) behavior at lower and intermediate values of wealth, $w < w_c$, and a power law (Pareto) tail at the larger wealth values, $w \geq w_c$, where $w_c$ is a crossover value that depends on the numerical fitting of the data. The Pareto law is expressed as:
\begin{equation}
f(w) \sim w^{-\alpha-1} \quad (w \geq w_c),
\label{pareto}
\end{equation}
where $f(w)$ is the probability density, and the exponent $\alpha$ is the \textit{Pareto exponent}, which has a value between 1 and 3. Reference~\onlinecite{Chakrabarti2013} gives concise accounts of empirical, numerical and analytical studies on this subject. Figure~\ref{fig_money2001} shows a plot of the cumulative probability distribution for wealth data. It is more practical to plot the cumulative probability distribution $C(w)=\!\int_w^\infty f(w')\,dw'$ as a
function of the wealth $w$, rather than $f(w)$, because statistical data is usually
reported at non-uniform intervals of $w$. Interestingly, when $f(w)$
is an exponential or a power-law function, then the respective $C(w)$ is also an
exponential or a power-law function. For more details of kinetic wealth exchange models, 
see Refs.~\onlinecite{Patriarca2010b,Chatterjee2007b,Chakraborti2011b}. 

Among the first examples of kinetic exchange models of markets proposed by non-physicists, we mention the work of Angle, a
sociologist.\cite{Angle1983a,Angle1986a,Angle2002a,Angle2006a}
In the first version of the model\cite{Angle1983a,Angle1986a} the
wealth exchanged at a encounter between two agents has the form of
Eq.~(\ref{deltax}), where $r_i = \epsilon \,\kappa \eta$ 
and $r_j = (1 - \epsilon) \,\kappa \eta$. Here 
$\kappa$ represents the maximum fraction of wealth that can be lost,
$\epsilon$ is a uniform random number in $(0,1)$, 
and $\eta$ is a stochastic variable with the values 0 or 1
with a probability distribution depending on the
difference between the wealth of the two agents $w_i - w_j$.
In the second model by Angle, referred to as the one-parameter
inequality process,\cite{Angle2002a,Angle2006a}
$r_i = \kappa \eta$ and $r_j = \kappa (1- \eta)$, 
with a similar meaning for $\kappa$ and $\eta$.
In contrast, in the basic version of the models\cite{Bennati1988a, Bennati1988b, Bennati1993a} introduced earlier by the economist,
Bennati,
$\Delta w_{ij}$ is a constant, independent of other parameters.

Independent of the above modeling efforts by social scientists, physicists also had made several studies. The first kind of models both with multiplicative and additive exchanges, were proposed by Ispolatov, Krapivski, and Redner.\cite{Ispolatov1998a}
In the kinetic exchange models introduced later, we first consider the work of Dra\-gu\-le\-scu 
and Ya\-ko\-ven\-ko,\cite{Dragulescu2000a}
where $r_i = 1 - r_j = \epsilon$, corresponding to a random reshuffling
scheme of the total wealth of the two agents.
We next consider a simple prototypical kinetic exchange model
with a $\Gamma$-function equilibrium distribution, which 
was introduced by Chak\-ra\-bor\-ti and Chak\-ra\-bar\-ti\cite{Chakraborti2000a}
to describe the trade activity between $N$ entities (for example, agents, firms, and companies)
employing a saving criterion when carrying out their trades.
Trades are represented by pair-wise wealth exchanges with
the dynamics given by Eq.~(\ref{exchange}) with
\begin{equation} 
\label{dx_CC}
\Delta w_{ij} = \kappa (\overline{\epsilon} \, w_i - \epsilon w_j )
= (1 - \lambda) (\overline{\epsilon} w_i - \epsilon w_j) \, ,
\end{equation}
where $\overline\epsilon
= 1 - \epsilon$. The exchange parameter $\kappa \in (0, 1)$
(or the saving parameter $\lambda = 1 - \kappa$)
defines the maximum fraction of the wealth $w$ in the exchange process
(or the minimum fraction of $w$ preserved during the exchange).
The parameter $\kappa$ (or $\lambda$) also determines the time scale 
of the relaxation process as well as the mean value $\langle w
\rangle$ in equilibrium.\cite{Patriarca2007a}
The equilibrium wealth distribution of the system is well described by 
the $\Gamma$-distribution in Eq.~(\ref{gamma-a}), where a fit suggests the empirical formula
\begin{equation}
\label{n}
\alpha = \frac{1 + 2\lambda}{1 - \lambda} = \frac{3}{\kappa} - 2 \, ,
\end{equation}
which relates the shape parameter $\alpha$ to the saving parameter $\lambda$.
Note that for $\lambda \to 0$ (or $\kappa \to 1$), $\alpha \to 1$, corresponding to the exponential function. In all the various models we have mentioned, the equilibrium distribution is well fitted by a
$\Gamma$-distribution.
This shape of the equilibrium distribution shows good agreement
with empirical wealth distributions at small and intermediate
values\cite{Yakovenko2001,Yakovenko2009a} (see Fig.~\ref{fig_money2001}). 

\begin{figure}
\centering
\includegraphics[width=8.5cm]{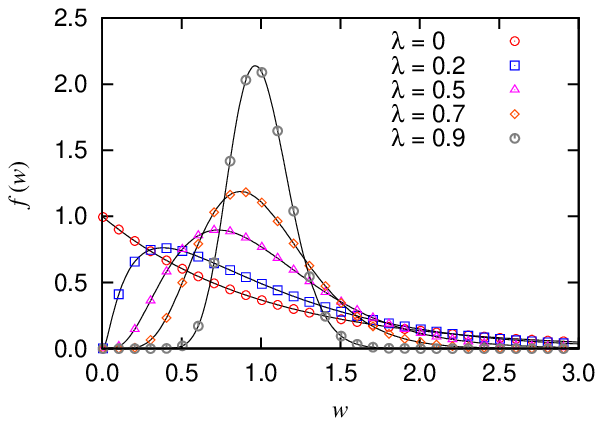}
\includegraphics[width=8.5cm]{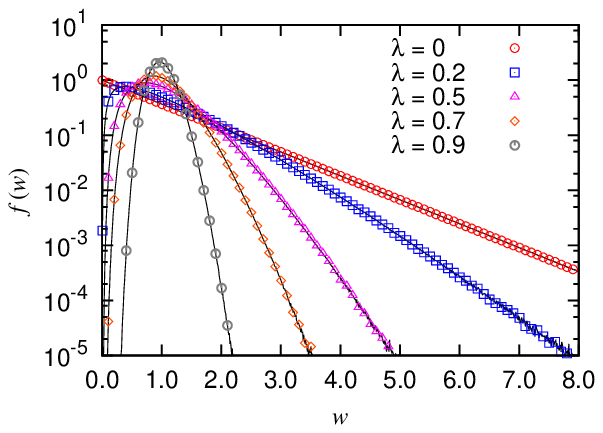}
\vspace{-0.5cm}
\caption{\label{fig_CC}
(Color online) Equilibrium wealth distributions of the model introduced in
Ref.~\onlinecite{Chakraborti2000a} and defined by Eq.~(\ref{dx_CC})
for different values of the saving parameter $\lambda$ for linear
(top) and semi-log (bottom) scales.}
\end{figure}

Numerical results of the model defined by
Eq.~(\ref{dx_CC}) are compared in Fig.~\ref{fig_CC} with 
a fit based on the $\Gamma$-distribution.
They were obtained using $10^4$ agents and 
$10^4$ time steps. The convergence to
equilibrium is fast and such a long simulation time was used only to
accumulate statistics (every 100 time steps). Note that here
one time step is actually a loop over $N$ exchanges. 
The number of agents employed has to be chosen sufficiently
large to ensure enough statistics --- implying a good quality
of the wealth histogram --- at the desired frequency scale, 
otherwise a very irregular histogram may appear at the smallest and
largest values of $w$.
The value $N = 10^5$ is still manageable computationally, yet
sufficiently large to clearly see frequencies as small as $10^{-5}$.

\subsection{\label{examples}Other social science applications}

Lallouache et al.\cite{Lallouache2010a} constructed a homogeneous model for the collective
dynamics of opinion formation in a society by modifying the kinetic
exchange dynamics studied in the context of markets. In this model
the opinion of the $i$th agent is represented by a continuous variable $w_i \in (-1,+1)$.
Interactions between agents who modify their opinions are assumed to
take place through random two-body encounters, with a 
dynamics defined by
\begin{subequations}
\begin{align}
\label{exchange1}
w_i & \to \lambda [ w_i + \epsilon w_j ] \\
w_j & \to \lambda [ w_j + \epsilon' w_i ] \, ,
\end{align}
\end{subequations}
where $\lambda \in (0,1)$ is the (constant) ``conviction'' parameter and
$\epsilon, \epsilon' \in (0,1)$ are uncorrelated uniformly distributed
stochastic processes. Note that there is no step-wise conservation of opinion, unlike for the preceding wealth models.
Remarkably, it is found that there is an appearance of polarity
or consensus, starting from initial random disorder (where the
$w_i$ are uniformly distributed with positive and negative values). 
In the language of physics, there is a ``spontaneous symmetry-breaking transition'' in the system: starting from the mean value of $w_i$, $\overline w(0)=0$, the system evolves either to the ``para'' state with $\overline w(t>\tau)=0$, where all agents have the opinion zero (for values of $\lambda \leq 2/3$), or to the ``symmetry broken'' state, with $\overline w(t>\tau)\neq 0$, where all the agents have either all positive or all negative opinions (for $\lambda \geq 2/3$). The time $t>\tau$, where $\tau$ is the relaxation time for the system. The relaxation behavior of the system shows a critical divergence of $\tau$ at $\lambda=\lambda_c=2/3$. 
Sen\cite{Sen2011a} generalized the model by Lallouache et al.\ by
introducing an additional parameter $\mu$ to represent the influencing
ability of individuals and studied the corresponding phase transitions.

In a different context, similar in spirit to the work presented in 
Refs.~\onlinecite{Iglesias2010a,Pianegonda2004a},
Ghosh et al.\cite{Ghosh2011a} considered an economic model in
which there is a poverty threshold $\theta > 0$ such that 
at any time $t$, at least one of the two interacting agents is
``poor;'' that is, its wealth satisfies $w < \theta$. 
The central role assigned to the poor traders produces
various new features such as a different form of the
equilibrium distribution and a phase transition in
the fraction of poor agents as a function of $\theta$. These models of kinetic exchanges, which draw inspiration from various socio-economic contexts, may also serve the purpose of introducing the ideas of phase transitions and critical phenomena in statistical physics.

\section{\label{heterogeneous}Heterogeneous Kinetic Exchange Models}

An interesting generalization of the homogeneous kinetic exchange
models we have discussed so far is the introduction of heterogeneity.
Probably the most relevant applications of heterogeneous kinetic
exchange models in the social sciences is the prediction of a
realistic shape for the wealth distribution which includes the Pareto
power law at the largest wealth values (see Fig.~\ref{fig_money2001}
and Sec.~\ref{social}).

\begin{figure}[t]
\centering
\subfigure[]{\includegraphics[width=8.5cm]{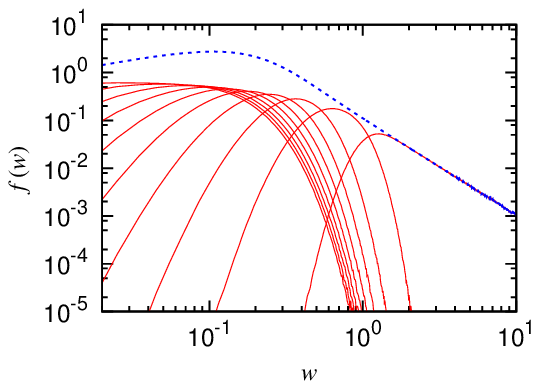}}
\hspace{1cm}
\subfigure[]{
\includegraphics[width=8.5cm]{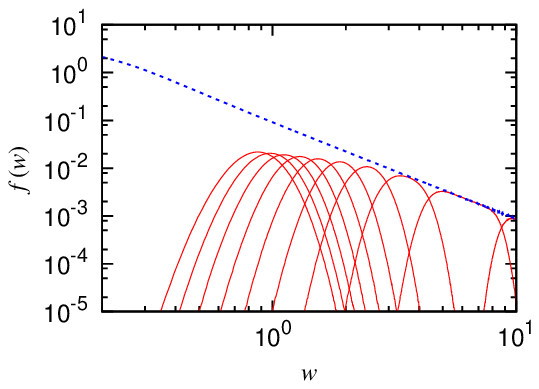}}
\caption{(Color online) Wealth distribution $f(w)$ for uniformly distributed $\kappa_i$ (or $\lambda_i$) in the interval (0,1); $f(w)$ is decomposed into partial distributions $f_i(w)$, where each $f_i(w)$ is obtained by counting the statistics of those agents with parameter $\lambda_i$ in a specific sub-interval (from Ref.~\onlinecite{Patriarca2006c}). (a) Decomposition of $f(w)$ into ten partial distributions in the $\lambda$-subintervals (0, 0.1), (0.1, 0.2) \dots (0.9, 1). (b) The last distribution of (a) in the $\lambda$-interval (0.9, 1) is decomposed into partial distributions obtained by counting the statistics of agents with $\lambda$-subintervals (0.9, 0.91), (0.91, 0.92) \dots (0.99, 1). Note how the power law appears as a consequence of the superposition of the partial distributions.
\label{fig_PARTIAL_A_NEW}}
\end{figure}

\begin{figure}[t]
\centering
\includegraphics[width=8.5cm]{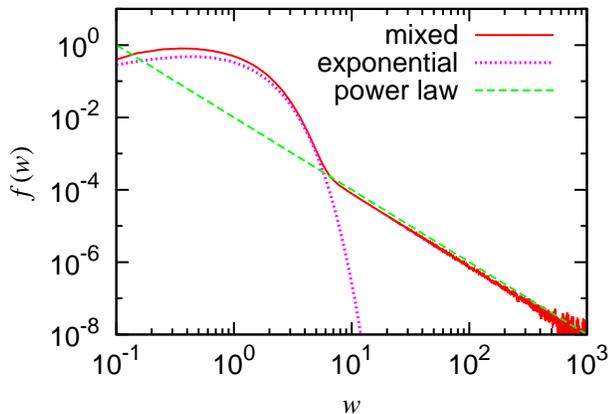}
\vspace{-0.5cm}
\caption{(Color online) Example of a realistic wealth distribution, from Ref.~\onlinecite{Patriarca2006c}.
Continuous curve: wealth distribution obtained by simulations of a mixed population of agents, such that 1\% of the agents have uniformly distributed saving propensities $\lambda_i \in (0, 1)$ and the other 99\% of the agents have $\lambda_i = 0.2$.
Dotted curve: exponential wealth distribution $\propto \exp(-w/\langle w \rangle)$, with the same average wealth, plotted for comparison with the distribution in the intermediate-wealth region.
Dashed curve: Pareto power law $\propto w^{-2}$, plotted for comparison with the large-income part of the distribution.
\label{fig_MIXED2_L0}}
\end{figure}

We consider again the model defined by Eq.~(\ref{dx_CC}) and introduce
heterogeneity by diversifying the
parameter $\kappa$ or equivalently the saving parameter
$\lambda$, meaning that each term $\kappa w_i$ (or $\lambda w_i$) is replaced by $\kappa_i w_i$ (or
$\lambda_i w_i$), 
thus obtaining\cite{Chatterjee2004}
\begin{equation}
\label{dx_CC1}
\Delta w_{ij} 
= \overline{\epsilon} \, \kappa_i w_i - \epsilon \kappa_j w_j
= \overline{\epsilon} (1 - \lambda_i) w_i - \epsilon (1 - \lambda_j) w_j
\, .
\end{equation}
As a simple example, we consider a set of heterogeneous agents
with parameters $\kappa_i$ uniformly distributed in the interval $(0,1)$.
By repeating the simulations using Eq.~(\ref{dx_CC1}), it is found that the
shape of the separate equilibrium wealth distributions $f_i(w)$ of each agent
is still a $\Gamma$-distribution. 
However, there is a surprise in the wealth distribution of the system $f(w)$,
given by the sum of the wealth distributions of the single agents,
$f(w) = \sum_i f_i(w)$.
As other analytical and numerical studies have also shown, 
$f(w)$ has an exponential form until intermediate $w$-values, and a Pareto power law develops
at the largest values of $w$ (see Fig.~\ref{fig_PARTIAL_A_NEW}).\cite{Chakraborti2009a,Patriarca2005a}
Such a shape is similar to real wealth distributions such as that shown in 
Fig.~\ref{fig_money2001}.
This shape of the equilibrium wealth distribution $f(w)$ is robust
with respect to the details of the system and the values of the other parameters, as
long as the values of the $\kappa_i$ are sufficiently spread over the
entire interval $\kappa = (0,1)$.
It is the group of agents with $\kappa \approx 0$ ($\lambda
\approx 1$) that are crucial for the appearance of a power law.
Not all agents have to differ from each other, as is best illustrated by repeating the simulation using a different
distribution for the $\kappa$-parameters or the $\lambda$s, in which
99\% of the agent population has $\lambda = 0.2$, and only
1\% of the population is heterogeneous with $\lambda$ 
in the interval $(0,1)$ (see Fig.~\ref{fig_MIXED2_L0}).

The heterogeneous model necessarily uses a finite upper cutoff $\lambda_{\max} < 1$,
when considering the saving parameter distribution, which directly
determines the cutoff $w_{\max}$ of the wealth
distribution, analogous to the cutoff observed in real distributions:
the closer $\lambda_{\max}$ is to one,
the larger $w_{\max}$ and the wider the interval in which 
the power law is observed.\cite{Patriarca2006c}

The $\lambda$-cutoff is closely related to the relaxation process,
whose time scales for agent $i$ is proportional to $1/(1 -
\lambda_i)$.\cite{Patriarca2007a}
Thus the slowest convergence rate is determined by $1 - \lambda_{\max}$.
The finite $\lambda$-cutoff used in simulations of heterogeneous kinetic exchange models
is not a limitation of the model, but reflects an important feature of real wealth distributions.

\section{Comments}

There are several analytical studies which complement the simulations and empirical studies. We refer the reader to Ref.~\onlinecite{Chakrabarti2013}, which gives concise accounts of studies on this subject, along with an exhaustive list of references.

The dynamics of kinetic exchange models are
sometimes criticized for being based on an approach that is
far from an actual economics perspective.
However, such a dynamics can also be derived from microeconomics theory.\cite{Chakrabarti2009a}
Although standard economics theory assumes that the activities of individual
agents are driven by the \textit{utility maximization principle}, 
the alternative picture that we have described is that the agents can be
viewed as particles exchanging ``wealth,'' instead of
energy, and trading in wealth (energy) conserving two-body scattering,
as in entropy maximization based on the kinetic theory of
gases.\cite{Chakraborti2009a} 
This qualitative analogy between the two maximization principles is
not new -- both economists and physicists had noted it in many
contexts, but this equivalence has gained firmer ground only
recently.\cite{Chakrabarti2009a}

\section{Suggested Problems}

The wealth exchange models, we have discussed are minimal and lend
themselves to various generalizations and modifications to study the
effect of additional features. 
Choose the kinetic exchange model, for instance, defined by:
$w_i \to w_i - \Delta w_{ij}$ and
$w_j \to w_j + \Delta w_{ij}$, with $\Delta w_{ij} = (1 - \epsilon)w_i - \epsilon w_j$. Perform the simulations and check whether the corresponding equilibrium Boltzmann-Gibbs distribution $f(w) = \beta \exp(-\beta w)$, where $\beta = 1/\langle w\rangle$ is reached, starting from any arbitrary initial distribution of $w$ amongst the agents.
Next, 
perform the simulations for $w_i \to w_i - \Delta w_{ij}$ and
$w_j \to w_j + \Delta w_{ij}$, when the form of $\Delta w_{ij}$ or the dynamical equations are modified, as suggested in the following.

\begin{itemize}
\item {\bf Different exchange rules}:
Verify that the final equilibrium wealth distribution remains Boltzmann-Gibbs distribution, when employing the rule that at each step an arbitrary constant amount [$\Delta w_{ij} = w_0$] is exchanged, or at each step a fraction of the average wealth in the system is exchanged.\cite{Dragulescu2000a} However, if instead the minimum wealth between the two agents is used as the exchanged amount [$\Delta w_{ij} = \min(w_i, w_j)$], condensation of the total wealth in the hands of a single agent will take place.\cite{Chakraborti2002a}

\item {\bf Taxation on transactions}:
Modify the dynamical equations by adding a term that takes as tax, a fixed amount of wealth $\delta w_0$ or a fixed percentage of the wealth exchanged $\Delta w_{ij}$ from agents $i$ and $j$, and redistributes it uniformly among all the others.\cite{Dragulescu2000a}

\item {\bf Debt}:
Modify the dynamical equations such that the agents are allowed to borrow or take loans up to a maximum amount $w_\mathrm{max}$ so that their wealth may become negative but not smaller than $-w_\mathrm{max}$. This modification can be implemented by increasing the total wealth by an additional amount $w_\mathrm{max}$.\cite{Dragulescu2000a}

\item {\bf Fixed saving}:
Add to the model a fixed saving $w_0$ for each of the two agents, so that the total wealth is diminished by an amount $2w_0$.\cite{Chakraborti2002a}

\item {\bf Many-agent interactions}: 
Set up a new dynamical rule in which wealth is redistributed in encounters between more than two agents, for example, among three agents $i$, $j$, and $k$. The other details of the transaction should be similar to that of a pair-wise transaction between agents $i$ and $j$.
Some care has to be taken to properly define the respective random fractions of wealth assigned to each agent to make the process symmetrical with respect to all the agents involved in the encounter.

\end{itemize}

\vspace{-0.5cm}
\begin{acknowledgments}
\vspace{-0.25cm}
We are grateful to all our collaborators and students. M.P. acknowledges financial support by the targeted financing project SF0690030s09 and by the Estonian Science Foundation through grant no. 9462.
\end{acknowledgments}

\end{document}